\documentclass[aps,pre,floats,floatfix,twocolumn]{revtex4}

\usepackage{ifthen}
\usepackage{ifpdf}
\usepackage{color}

\ifpdf
\usepackage{graphicx}
\usepackage{epstopdf}
\else
\usepackage{graphicx}
\usepackage{epsfig}
\fi

\usepackage{latexsym}
\usepackage{amsmath}
\usepackage{amssymb}
\usepackage{bm}
\usepackage{wasysym}

\newcommand{\eexp}{\mbox{e}^}

\newcommand{\mass}{\mathsf{m}}

\newcommand{\tbox}[1]{\mbox{\tiny #1}}
 
\newcommand{\mylabel}[1]{\label{#1}} 
\newcommand{\beq}{\begin{eqnarray}}
\newcommand{\eeq}{\end{eqnarray}} 
\newcommand{\be}[1]{\begin{eqnarray}\ifthenelse{#1=-1}
{\nonumber}{\ifthenelse{#1=0}{}{\mylabel{e#1}}}}
\newcommand{\ee}{\end{eqnarray}} 

\newcommand{\Eq}[1]{\textcolor{blue}{Eq.\!\!~(\ref{#1})}} 
\newcommand{\Fig}[1]{\textcolor{blue}{Fig.}\!\!~\ref{#1}} 
\newcommand{\hide}[1]{}
\renewcommand{\cite}[1]{\textcolor{blue}{[\onlinecite{#1}}]} %{[\onlinecite{#1}]} 

\begin{document} 

\title{Straightforward quantum-mechanical derivation of the \\ Crooks fluctuation theorem and the Jarzynski equality}

\author{Doron Cohen$^1$ and Yoseph Imry$^2$} 

\affiliation{
$^1$\mbox{Department of  Physics,  Ben Gurion University of the Negev, Beer Sheva 84105, Israel} 
\\
$^2$\mbox{Department of Condensed Matter Physics, Weizmann Institute of Science, Rehovot 76100, Israel}
}

% \date{\today}

\begin{abstract}
We obtain the Crooks and the Jarzynski non-equilibrium fluctuation relations using a direct quantum-mechanical 
approach for a finite system that is either isolated or coupled not too strongly to a heat bath. 
These results were hitherto derived mostly in the classical limit. The two main ingredients 
in the picture are the time-reversal symmetry and the application of the first law to the case 
where an {\em agent} performs work on the system. No further assumptions regarding stochastic or Markovian 
behavior are necessary, neither a master equation or a classical phase-space picture are required. 
The simplicity and the generality of these non-equilibrium relations are demonstrated, giving 
very simple insights into the Physics.
\end{abstract}



\maketitle

%%%%%%%%%%%%%%%%%%%%%%%%%%%%%%%%%%%%%%%%%%%%%%%%%%%%%%%%%%%%%%%%
\section{Introduction}

In contrast to the situation in equilibrium statistical physics, and linear response theory, 
there are not so many well-established results for systems far from equilibrium \cite{NFT1,NFT2,NFT3}. 
Two such extremely interesting results are the ``nonequilibrium fluctuation  theorem" (NFT) 
of Crooks \cite{Crooks} and the related Jarzynski equality \cite{Jarzynski,VB}. 
Both have to do with the work done by/on a finite system coupled to a heat bath. 
We also mention here previous works \cite{rsp,Gavish,book}, 
showing that the Kubo formalism, the fluctuation-dissipation theorem, 
and the associated detailed-balance relations are valid 
in a large class of nonequilibrium steady-state systems, 
and not only in equilibrium.

The system under consideration is described by a time dependent Hamiltonian ${\cal H}(X(t))$, 
where the parameter $X$ is a time-dependent c-number, often coupled linearly to an observable of the system.
At ${t=t_0}$ the system is prepared in thermal equilibrium at the temperature~$T$. 
The thermalization is achieved by connecting it for a long enough time to a thermal bath 
at that temperature. After that, within the time ${t=t_1}$, the system undergoes a ``work process".
This means that an ``agent" changes the value of~$X$ from~$X_0$ to~$X_1$.
During this process a work $\mathcal{W}$ is performed on the system, 
and possibly some heat $\mathcal{Q}$  is dissipated into the bath \cite {LL}. 
In the simplest scenario the system is isolated, and heat flow is not involved.  
It should be emphasized that at the end of the work process, the system is in general {\em not} in equilibrium.
 
The NFT deals with the probability distribution $P(\mathcal{W})$ 
of the work $\mathcal{W}$, whose experimental determination requires 
to repeat the process protocol many times, and to record the 
measured values of $\mathcal{W}$. 
Specifically the NFT concerns the ratio $P_{0\leadsto1}(\mathcal{W})/P_{1\leadsto0}(-\mathcal{W})$
between the statistics of the forward scenario, and the statistics of the reversed scenario.      
In the latter, the system is equilibrated with the same bath 
under conditions such that ${X=X_1}$. Then the time-reversed process protocol 
is realized, such that at the final time ${X=X_0}$.  

Derivations of various quantum mechanical versions of the NFT 
have been discussed in several publications \cite{q1,q2,q3,NFTx,hanggi,CrooksJQ,VV}. 
However, a major subtlety arises with regard to the definition 
of work. Citing the introduction of Ref.\cite{CrooksJQ}: 
``The generalization of the Jarzynski identity to closed system quantum dynamics is
technically straightforward [...] However, for a system that can interact with
the environment this does not suffice [...] Unlike for a classical system, we cannot
continuously measure the energy of the system without severely disturbing the dynamics
of the system".

The objective of our paper is to present a simple derivation of the NFT 
in the quantum mechanical context, bypassing various subtleties that, 
in our view, have obscured the simple physics involved. 
The main issue is to define carefully the notion of work in the quantum mechanical context, 
and to clarify the role that is played by the bath.      

%%%%%%%%%%%%%%%%%%%%%%%%%%%%%%%%%%%%%%%%%%%%%%%%%%%%%%%%%%%%%%%%%%%%%%%%%
{\em Outline: } 
We refer to the evolution during a work process, 
and formulate for it a generalized detailed 
balance relation. Then we discuss the notion of work,  
leading to the NFT of Crooks. The main issue is 
the modeling the work agent, and the understanding
of the role that is played by the bath.
The implied Jarzynski equality and the implications
on the dissipated work and on the entropy production
are briefly discussed. 

%%%%%%%%%%%%%%%%%%%%%%%%%%%%%%%%%%%%%%%%%%%%%%%%%%%%%%%%%%%%%%%%
\section{Evolution during a work process}

The system under consideration is described by a time-dependent Hamiltonian ${\cal H}(X(t))$.
Let us assume that a classical ``agent" changes the value of the c-number control parameter~$X$ 
from~$X_0$ at ${t=t_0}$ to~$X_1$ at ${t=t_1}$. In some cases, but not in general, 
the actual duration of the time dependent stage might be ${\tau\ll|t_1-t_0|}$.  
Given that at ${t=t_0}$ the system has been prepared in some eigenstate $n^{(0)}$ of ${\cal H}(X_0)$, 
we ask what is the probability $\mathrm{P}_{0\leadsto1}(m|n)$  that 
at the later time ${t=t_1}$ it is measured in an eigenstate $m^{(1)}$ of ${\cal H}(X_1)$.
Below we use the notation 
\beq
\omega \ \ = \ \ E_m^{(1)} - E_n^{(0)}
\eeq  
In a later section we shall define the notion of work~$\mathcal{W}$,   
and shall explain that up to some uncertainty,  
we can make the identification ${\mathcal{W}=\omega}$, 
provided the system is isolated from the environment.

For a strict quantum adiabatic process   
one has $\mathrm{P}(m|n)=\delta_{n,m}$. 
But we are interested in more general 
circumstances. In particular we focus 
in this section on unitary evolution 
for which 
\be{2}
\mathrm{P}_{0\leadsto1}(m|n) \ \ = \ \  
\Big|\langle m^{(1)}|U_{0\leadsto1}| n^{(0)} \rangle\Big|^2 , 
\eeq
where $U$ is the time-evolution operator.
What is important for the derivation of the NFT 
is the micro-reversibility of the dynamics, namely, 
\be{3} 
\mathrm{P}_{1\leadsto0}(n|m) \ \ = \ \  \mathrm{P}_{0\leadsto1}(m|n)
\eeq
Note that in general the reversed process requires 
to transform some fields, e.g. to change the sign of 
the magnetic field if present. 

For completeness it is also useful to define the notion of ``classical dynamics".
Given phase-space, we can divide it into cells with some arbitrary desired resolution. 
Then we can regard~$n$ as an index that labels cells in phase space. 
The classical equations define a map 
\beq 
|n^{(1)}_{final}\rangle \ \ =  \ \ \mathsf{M} \ |n^{(0)}_{initial}\rangle
\eeq
We use quantum style notations in order to make the relation
to the quantum formulation clear. It follows that $\mathrm{P}(m|n)$, 
instead of being a stochastic kernel, becomes a deterministic 
kernel that induces {\em permutations}
\beq  
\mathrm{P}_{0\leadsto1}(m|n) \Big|_{classical} \ \ = \ \ \delta_{m,\mathsf{M}n}
\eeq
The derivation in the next section does not depend on 
whether the dynamics is ``classical" or ``stochastic" or ``quantum" 
in nature as long as the {\em measure} and the micro-reversibility 
are preserved. The preservation of {\em measure} is reflected 
by our discrete notations: If, say, we had deterministic dynamics 
that does not satisfy Liouville theorem, we could not have used  
the above ``cell construction"

%%%%%%%%%%%%%%%%%%%%%%%%%%%%%%%%%%%%%%%%%%%%%%%%%%%%%%%%%%%%%%%%
\section{The generalized detailed balance relation}

The power spectrum of the fluctuations of an observable~$A$ 
is given by the following spectral decomposition  
\beq
\tilde{S}(\omega) \ \ = \ \ 
\sum_{n,m} p_n
\ \Big|\langle m | A | n \rangle\Big|^2  
\ \delta\Big(\omega-(E_m-E_n)\Big) 
\eeq
Here we assume a time independent Hamiltonian, 
and stationary preparation that can be regarded   
as a mixture of eigenstates with weights~$p_n$. 
For a canonical preparation  
\beq
p_n \ \ = \ \ \frac{1}{Z} \eexp{-E_n/T} \ \ = \ \ \exp\left[-\frac{E_n-F(X_0)}{T}\right]
\eeq
where $Z$ is the partition function, 
and $F(X)$ is the Helmholtz free energy 
at temperature~$T$, calculated here 
for the fixed value of the control parameter~$X$. 
Then one obtains after two lines of straightforward 
algebra, the detailed balance relation 
\be{8}
\frac{\tilde{S}(\omega)}{\tilde{S}(-\omega)} 
\ \ = \ \ \exp\left[ \frac{\omega}{T} \right]
\eeq
This relation plays a key role in the linear response 
theory. Specifically it reflects the ratio between 
the tendency of the system to absorb and emit energy 
from/to a driving source~$-f(t)A$.

In complete analogy we define the following spectral kernel: 
\beq 
&& P_{0\leadsto1}(\omega) 
\\ \nonumber
&& \ \ = \sum_{n,m} p_n^{(0)}
\ \mathrm{P}_{0\leadsto1}(m|n) 
\ \delta\Big(\omega-(E_m^{(1)}-E_n^{(0)})\Big) 
\eeq
Here the superscript indicates whether we refer to 
the initial Hamiltonian $\mathcal{H}(X_0)$ 
or to the final Hamiltonian $\mathcal{H}(X_1)$.
For the reversed process we write  
\beq 
&& P_{1\leadsto0}(\omega) 
\\ \nonumber
&& \ \ =  \sum_{m,n} p_m^{(1)}
\ \mathrm{P}_{1\leadsto0}(n|m) 
\ \delta\Big(\omega-(E_n^{(1)}-E_m^{(0)})\Big) 
\eeq
It immediately follows, in analogy with the usual detailed balance condition, 
that the ratio of the spectral functions $P_{0\leadsto1}(\omega)$ and $P_{1\leadsto0}(-\omega)$  
is determined by the ratio of the initial probabilities $p_n^{(0)}$ and  $p_n^{(1)}$, 
leading to 
\begin{equation} \label{CrooksE}
\frac{P_{0\leadsto1}(\omega)}{P_{1\leadsto0}(-\omega)} \ \ = \ \ \exp\left[\frac{\omega - (F(X_1)-F(X_0))}{T}\right],
\end{equation}
where both $F(X_1)$ and $F(X_0)$ refer to the same 
preparation temperature~$T$. Note again that if~$X$ 
does not change in time, this relation formally coincides 
with the detailed balance relation \Eq{e8}.

%%%%%%%%%%%%%%%%%%%%%%%%%%%%%%%%%%%%%%%%%%%%%%%%%%%%%%%%%%%%%%%%
\section{The notion of work and the NFT of Crooks}

The main difficulty in the quantum formulation of the NFT concerns 
the definition of work \cite{hanggi,CrooksJQ,w0,w1,w2}. 
Consider first an isolated system. 
Naively we can define ${\mathcal{W}=\omega}$, namely the 
work is the change in the energy of the system.
But in the quantum reality this means that we have to do 
a measurement of the initial stage, hence the state 
of the system collapses and it is no longer canonical.

Furthermore, assume that we want to consider a multi-stage 
process that extends over two time intervals ${t_0\rightarrow t_1\rightarrow t_2}$.  
We would like to say that the work done is the sum $\mathcal{W}_{0\leadsto1} + \mathcal{W}_{1\leadsto2}$.
With the above definition we have to perform 
a measurement at the time instant $t_1$. But 
in the quantum mechanical reality we might not have 
the time for that (see further discussion below).

It is therefore clear that the definition of work 
requires refinement. One possible direction \cite{hanggi} is to define
\beq
\mathcal{W}_{0\leadsto1} = \int_{t_0}^{t_1}  \frac{\partial \mathcal{H}}{\partial X} \dot{X} dt
\eeq
Then, in analogy with the theory of {\em counting statistics} \cite{levitov,nazarov,cnz},
one might say that a {\em continuous measurement} 
is required, involving a weak coupling to a von Neumann pointer.
The problem with this approach is that the  
counting statistics {\em quasi probability} \cite{nazarov,cnz} 
has no simple physical interpretation, and might be even negative.

It turns out that in the present context there is a simple 
way out of these subtleties, that parallels the 
conventional classical perspective \cite{w0}. 
Instead of regarding $X(t)$ as a c-number field, 
we regard it as a dynamical variable 
of an ``agent" that is doing work. The total 
Hamiltonian can be formally written as  
\be{122}
\mathcal{H}_{\tbox{total}} \ \ = \ \ \mathcal{H}(r;X) +  \mathcal{H}_{agent}(X)
\eeq
where $r$ stands for system dynamical variables, 
and $X$ is the agent degree of freedom.  
Then we define the work $\mathcal{W}$ 
as the change in the energy of the agent
\beq
\mathcal{W} \ \ = \ \ E_{agent}(0) - \mathcal{H}_{agent}
\eeq
where $E_{agent}(0)$ is its initial energy, 
which is assumed to be well defined up to some 
small uncertainty.

It should be clear that by treating the agent 
as a dynamical variable we bypass the energy-time uncertainty 
fallacy, as discussed long ago~\cite{AB}.
Once the energy is transferred to an agent, 
there is no theoretical limitation on the accuracy of its measurement, 
irrespective of the time-of-measurement issue.

In our treatment it is assumed that we have control 
over the strength of the interaction between the system 
and the agent. Hence we have the option to switch ``on" 
the interaction in two possible ways: 
{\bf \ (i)} within a restricted region in ${(r,X)}$ space;  
{\bf \ (ii)} within a restricted time duration. 
In the latter case the Hamiltonian \Eq{e122} would become 
time dependent, and consequently we would not have control 
over the precise $X$~displacement of the agent.
    
Once the notion of work is clarified  
it follows automatically that for an isolated 
system $\mathcal{W}=\omega$, to the extent that 
the unavoidable quantum uncertainties can be ignored.
From here follows the Crooks relation 
\begin{equation}  \label{Crookswork}
\frac{P_{0\leadsto1}(\mathcal{W})}{P_{1\leadsto0}(-\mathcal{W})} 
\ \ = \ \ \exp\left[\frac{\mathcal{W} - (F(X_1)-F(X_0))}{T}\right]
\end{equation}
Instead of going on with an abstract discussion of 
what do we mean by ``work agent", we consider below
two simple prototype models that illuminate this notion.

%%%%%%%%%%%%%%%%%%%%%%%%%%%
\begin{figure}%[h!]

\includegraphics[clip,width=0.8\hsize]{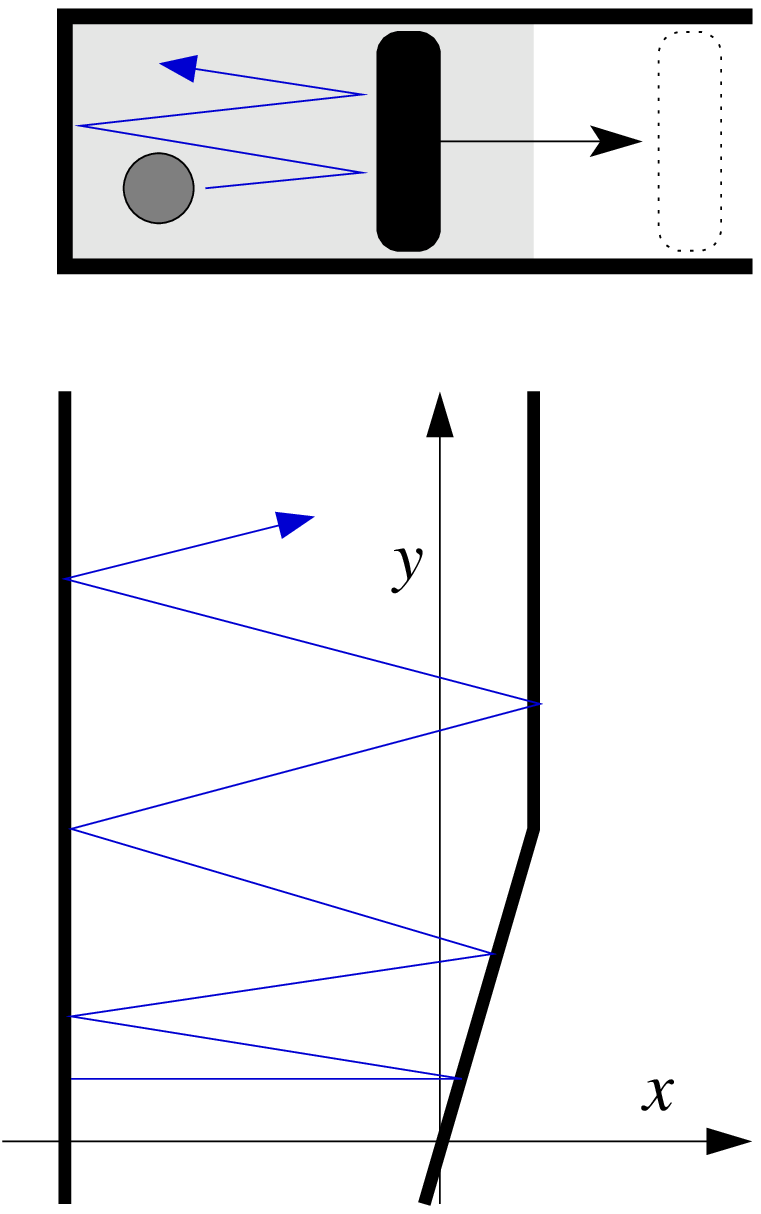}

\caption{
The system is a gas of particles in a box. The box region is indicated by grey.
A representative trajectory is illustrated. The ``agent" on which work is being done 
is a piston that is free to move to infinity. After the piston is pushed out the gas 
particles stay in the box, and no longer interact with the piston, but possibly 
may interact (say) with a bath or with other agents. At the end of each single "run" 
of the experiment, there is an unlimited time to measure the energy of the freely 
moving piston in the desired resolution.}

\label{f1}
\end{figure}
%%%%%%%%%%%%%%%%%%%%%%%%%%%

%%%%%%%%%%%%%%%%%%%%%%%%%%%%%%%%%%%%%%%%%%%%%%%%%%%%%%%%%%%%%%%%
%%%%%%%%%%%%%%%%%%%%%%%%%%%%%%%%%%%%%%%%%%%%%%%%%%%%%%%%%%%%%%%%
\section{Modeling the work agent}

In order to define the notion of work we find it essential 
to regard the ``agent" as a dynamical entity. It can be 
another object (``piston") from/to which energy is transferred,  
or it can be a field with which the system interacts, absorbing  
or emitting excitations (``photons").

%%%%%%%%%%%%%%%%%%%%%%%%%%%%%%%%%%%%%%%%%%%%%%
\subsection{Modeling the agent as a piston}

\begin{figure}%[h!]

\includegraphics[clip,width=0.8\hsize]{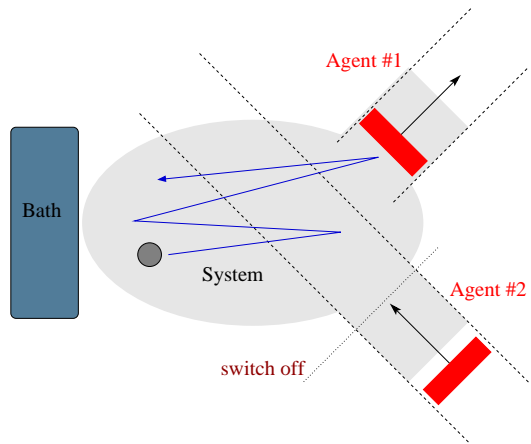}

\caption{(color online)
Illustration of a multi-stage process that consists 
of time intervals during which the system interacts 
with a bath, and with two different agents.
The system is a gas particles 
that are confined to move in the shaded area.  
The interaction with the first agent is as described in \Fig{f1}. 
The second agent compresses that gas, until the interaction with 
it is switched off: then it becomes like a free particle.}

\label{f2}
\end{figure}

The prototype model for explaining the notion of work 
in standard thermodynamics textbooks is the gas-piston 
system that is illustrated in \Fig{f1}. 
The ``agent" on which work is being done 
is a piston that is free to move to infinity. 
After the piston is pushed out the gas 
particles stay in the box, and no longer interact 
with the piston, but possibly may interact (say) 
with a bath or with other agents as in \Fig{f2}. 
At the end of each single "run" of the experiment, 
there is an unlimited time to measure the energy of the freely 
moving piston in the desired resolution.

The essential ingredient in the illustrated construction 
is the decoupling at the end of the interaction:  
After the piston moves outside of the shaded region, 
it becomes a free object whose kinetic energy we can measured
without having a time limitation. 

For presentation purposes, but without any loss 
of generality, we consider a single gas particle       
and regard the box as one dimensional. 
The Hamiltonian is 
\beq
\mathcal{H}_{\tbox{total}}(r,p;X,P) = 
\frac{p^2}{2\mass}+V_{\tbox{box}}(r) 
+ u(r{-}X)+\frac{P^2}{2M}
\eeq
where $u(r-X)=u_0\delta(r-X)$ with ${u_0=\infty}$.
Thanks to a potential $V_{\tbox{box}}$ the gas particle  
remains in the shaded region even if the piston is ``out". 
Once the piston is out the ``system" no longer 
affects the ``agent", nor affected by it. 

In order to visualize the dynamics it is convenient 
to define $\alpha=(\mass/M)^{1/2}$, and ${p_x=p}$, 
and ${p_y=\alpha P}$, and ${x=r}$, and ${y=(1/\alpha)X}$.
Then the Hamiltonian takes the form
\beq
\mathcal{H}_{\tbox{total}} = 
\frac{1}{2\mass}(p_x^2+p_y^2) + V_{\tbox{box}}(x) + u(x-\alpha y) 
\eeq
We assume that initially the piston is prepared in rest 
with some uncertainty $\Delta X$ in its position, 
and an associated uncertainty ${\Delta P \sim 1/\Delta X}$ 
in its momentum. Accordingly the uncertainty of the total 
energy is 
\beq
\Delta E \ \ \sim \ \ [M \Delta X^2]^{-1} + \Delta E_{\tbox{system}}
\eeq
The total energy~$E$ is a constant of motion.
It follows that the probability distribution 
of the total energy is a $\delta$~function.
The total energy~$E$ is the sum of the particle 
energy and the piston energy. Let us denote 
the increase in the particle energy as~$\omega$, 
and the decrease in the piston energy as~$\mathcal{W}$.
It follows that the joint distribution is 
\beq 
\mathrm{P}(\omega,\mathcal{W}) \ \ = \ \ P(\omega) \ \delta(\mathcal{W}-\omega) 
\eeq
where the equality is justified to the extent 
that $\Delta E$ can be neglected. 
Under such conditions the distribution of work $P(\mathcal{W})$ 
is the same as $P(\omega)$.

The argument above has established the equality  
of $P(\mathcal{W})$ and $P(\omega)$  
for a system that is prepared in a microcanonical state, 
such that $\Delta E_{\tbox{system}}$ is a small uncertainty.
But trivially the equality of the two distributions 
extends to any mixture, and in particular to the canonical 
preparation under consideration. 
We note that our definition of $P(\omega)$   
in the previous section has assumed a c-number driving source, 
while here there is some uncertainty $\Delta X$ in the position      
of the piston. Accordingly a trade-off is required 
with regard to $\Delta X$ and $\Delta E$. This 
trade-off is the physical limit of the NFT applicability.  
In practice, and in particular for large deviations, 
this uncertainty should not be an issue.

%%%%%%%%%%%%%%%%%%%%%%%%%%%
%%%%%%%%%%%%%%%%%%%%%%%%%%%

%%%%%%%%%%%%%%%%%%%%%%%%%%%%%%%%%%%%%%%%%%%%%%%%%%%%%% 
\subsection{Modeling the agent as a field}

In this subsection we consider another illuminating example 
for a ``work agent" but with a different emphasis: 
we would like to illuminate the role that is played 
by the strength of the system-agent interaction.
For this purpose the piston model is somewhat unnatural 
because perturbation theory is not well controlled. 
This is the motivation to consider a different example. 
Below the agent is a field with which the system interacts, 
and the measurement is the detection of a field quanta. 
These quanta can be observed at any later time 
without disturbing the on-going driving cycle. 

For sake of clarity the reasoning below is based on 
the traditional weak coupling assumption. Namely, 
we assume that the driving induces transitions that are 
determined by the Fermi-Golden-Rule.
While we employed below perturbation-theory thinking, 
we re-emphasize that these considerations 
are much more general: for stronger perturbations, 
one may think in terms of the evolution operator~$U$ 
of \Eq{e2}, and microscopic reversibility \Eq{e3} 
follows {\em mutatis mutandis}.

Consider a classical force $\mathcal{F}$ that arises, say, 
from  a classical electric field that acts on charged particles.
Taking the coupling to be via the total dipole moment of the system, 
the interaction term is 
\beq
\mathcal{H}_{\tbox{system-agent}} \ \ = \ \ - \mathcal{F}(t) \sum  r_i
\eeq
where the $r_i$ are the coordinates of the particles along the relevant axis, 
and  $\mathcal{F}(t)$ is a c-number force that is switched from ${\mathcal{F}_0=0}$ 
at $t_0=0$, to ${\mathcal{F}_1=\delta\mathcal{F}}$ at time $t_1$. 

To see what is going on, think of expanding $\mathcal{F}(t)$ in a Fourier integral.  
The Fourier components $\mathcal{F}_\omega$ are significant on an interval 
of order $1/\tau$, where $\tau$ is the actual duration of the variation,
which is possibly small compared with ${|t_1-t_0|}$.
Small $\delta\mathcal{F}$ and/or small $\tau$ 
make the relevant Fourier components small. 
From low order perturbation theory it follows that 
the transitions are to levels~$E_m$ whose energy 
is within $\sim 1/\tau$ of the initial energy~$E_n$, 
with probabilities proportional to $|\mathcal{F}_{\omega}|^2$.
Very importantly, energy is conserved in the sense that  
the excitation takes an energy ${\omega=E_m-E_n}$ 
from the field.
We know that if we quantize the field ${\mathcal{F}}$, a photon 
with the energy ${\omega=E_m-E_n}$ will be destroyed during the transition.

A side note is in order:
for a closed system, the work done by the classical agent 
is all converted to a change of the system energy. 
A well-known even stronger example is that of a probe particle 
inelastically scattered from the system losing an certain energy 
which is then equal to the energy of the created excitation(s).

In the absence of a coupling to the bath the transitions 
are into an energy range ${\Gamma_{F} \approx 1/\tau}$ that may contain 
many states. When a  coupling to the bath is introduced, the levels 
of the system acquire an additional width $\Gamma_{B}$. 
If the interaction is weak enough $\Gamma_{B}$ becomes 
smaller than the mean level spacing of the {\em system}.

Before going on with the above reasoning we would like 
to recall what is the justification for  the canonical state.
The reader is most probably familiar with the standard 
textbook argumentation in \cite{LL}: if a system is weakly 
coupled to a bath its energy distribution will approach 
a canonical distribution, as postulated by Gibbs, 
based on an ergodicity assumption. 
There is an interesting refined version of this argument 
that has been introduced by \cite{Lipkin}. Namely, one can 
rigorously show that the system would equilibrate to 
a canonical mixture, with zero off diagonal elements, 
if $\Gamma_{B}$ is smaller than the mean level spacing 
of the {\em system}. This weak coupling assumption is crucial 
whenever we try to connect statistical mechanics with 
thermodynamics, and in particular it is essential for the 
following argumentation.

Coming back to the work process scenario, 
it is clear that in order to relate the backward 
and the forward process we have to assume that the system starts 
in a canonical mixture state. If the system interacts 
with a bath it is essential to assume that in the 
preparation stage, either of the forward or of the reversed 
process, the system-bath coupling is small enough such 
that the system eigenstates are not mixed.
This is what counts in obtaining \Eq{CrooksE}.
Other than that, energy conservation implies 
that $\mathcal{W}=\omega$, so again, the 
distributions of $\omega$ and of $\mathcal{W}$ 
are the same, hence \Eq{Crookswork} follows.

%%%%%%%%%%%%%%%%%%%%%%%%%%%%%%%%%%%%%%%%%%%%%%%%%%%%%%%%%%%%%%%%%%%%%%%%%%%%%%%%%%%%%%%%%%%%%%
\section{The irrelevance of the bath}

The Crooks relation and the Jarzynski equality
concern the probability distribution of {\em work} done
during  a non-equilibrium process 
that starts with a canonical state.
We deduced in the previous sections that in the case 
of an isolated system $P(\mathcal{W})$ 
satisfies the same Crooks relation as $P(\omega)$. 
We now want to extend the validity of this relation to the 
case of non-isolated system.

It is clear that the bath is likely to affect 
significantly the dynamics. In some cases the dissipative 
dynamics can be described by a Markovian master equation - 
but we do not want to impose this assumption.  
Rather, as discussed in last part of section V, 
we are satisfied with the traditional assumption 
of small system-bath coupling: it is the same 
assumption that justifies the emergence of the 
canonical mixture upon preparation \cite{Lipkin}. 
Within the framework of this traditional assumption,    
let us discuss whether the interaction with
the bath can affect the Crooks relation.

{\bf First scenario.-- } 
After the work process has ended we allow the system 
to relax to the bath temperature~$T$.
This additional step does not involve work,
as noted in \cite{VB}, 
hence $P(\mathcal{W})$ is not affected.

{\bf Second scenario.-- } 
Assume that there is a finite system-bath coupling~$\eta$ during the process.
The duration of the process is~$\tau$.
Inspired by the argumentation of \cite{Jarzynski},
we regard the system and the bath as one grand-system, 
for which 
\beq\nonumber
&&\frac{P_{0\leadsto1}(\mathcal{W};\eta,\tau)}{P_{1\leadsto0}(-\mathcal{W};\eta,\tau)}
\\
&& \ \ \ =  \exp\left[ \frac{\mathcal{W} - (F_{tot}(X_1;\eta)-F_{tot}(X_0;\eta))}{T} \right]
\eeq
It should be clear that $P_{0\leadsto1}(\mathcal{W})$ and $P_{1\leadsto0}(-\mathcal{W})$ depend
on both $\eta$ and $\tau$. But the ratio, according to Crooks,
is independent of~$\tau$. Still one suspects that the right-hand side
depends on $\eta$. But in fact this is not so. The argument is as follows:
The ratio is independent of $\tau$, and therefore we can evaluate it,
without loss of generality, for $\tau\rightarrow0$;
But in this "sudden" limit the result should be independent of $\eta$,
because the bath has no time to influence the work.
We therefore can set ${\eta=0}$, and deduce that without loss of generality
\beq\nonumber
&&\frac{P_{0\leadsto1}(\mathcal{W};\eta,\tau)}{P_{1\leadsto0}(-\mathcal{W};\eta,\tau)}
\\
&& \ \ \ =\exp\left[ \frac{\mathcal{W} -(F_{sys}(X_1)-F_{sys}(X_0))}{T} \right]
\eeq
without dependence on $\eta$ and $\tau$. 
Hence the NFT is established for a process 
in which the system is non-isolated. 
In particular, it may interact with a thermal reservoir.

%%%%%%%%%%%%%%%%%%%%%%%%%%%%%%%%%%%%%%%%%%%%%%%%%%%%%%%%%%%%%%%%%%%%%%%%%%%%%%%%%%%%%%%%%%%%
\section{The Jarzynski equality}

It is well known \cite{Crooks} that the Jarzynski equality \cite{Jarzynski} 
is an immediate consequence that follows from the Crooks relation \Eq{Crookswork}.
For completeness we repeat this derivation here.
Multiplying both sides of the Crooks relation by $P_{1\leadsto0}(-\mathcal{W})$, 
integrating over $\mathcal{W}$, and taking into account the normalization 
of $P(\mathcal{W})$, one obtains
\beq
\left\langle \exp\left[-\frac{\mathcal{W}}{T}\right] \right\rangle 
\ \ = \ \ \exp\left[ - \frac{F(X_1)-F(X_0)}{T} \right],
\eeq
which is the Jarzynski relation. From here follows that 
\be{24}
\left\langle \mathcal{W} \right\rangle \ \ > \ \ F(X_1)-F(X_0)
\eeq
This variation of the 2nd law of thermodynamics  
is known as the {\em maximum work principle}, 
because it sets an upper bound on the work $W=-\mathcal{W}$ 
that can be {\it extracted} from a work process. 
Optionally it can be regarded as the minimum work $\mathcal{W}$
needed from the agent to do the process~\cite{LL}. 
Note that our sign conventions for $\mathcal{W}$ and $\mathcal{Q}$
are opposite to those that are used in most textbooks.

%%%%%%%%%%%%%%%%%%%%%%%%%%%%%%%%%%%%%%%%%%%%%%%%%%%%%%%%%%%%%%%%%%%%%%%%%%%%%%%%%%%%%%%%%%%%
\section{Dissipated work and entropy production}

It is instructive to recast the Crooks relation \Eq{Crookswork} in terms 
of entropy produced, as in fact was originally formulated by Crooks. 
From \Eq{e24} it follows that the difference ${\Delta F = F(X_1)-F(X_0)}$
is the minimum work that is required in a reversible quasi-static process. 
Accordingly the difference ${\mathcal{W}- \Delta F}$ 
can be regarded as the dissipated work in a realistic process.
Dividing by~$T$ we get a quantity $\Delta S_w$ that we regard 
as the entropy production. For the temperature we use units such that 
the Boltzmann constant is unity. Consequently the fluctuation theorem \Eq{Crookswork} reads: 
\begin{equation}  \label{Crooksentropy}
\frac{P_{0\leadsto1}(\Delta S_w)}{P_{1\leadsto0}(-\Delta S_w)} 
\ \ = \ \ \exp\left[\Delta S_w\right]
\end{equation}
Below we would like to better clarify the connection with thermodynamics, 
and in particular with the Clausius version of the 2nd law. 
   
Taking a puristic point of view, one defines thermodynamic functions 
only for equilibrium states. Therefore let us assume that the system 
ends up in a thermodynamic equilibrium, say by allowing 
it to relax at the end of the driving process. Under this assumption 
we can associate with the initial and final states well defined values 
of system entropy, whose difference can be expressed using thermodynamic 
functions:
\beq
\Delta S  \ \ = \ \ \frac{\Delta E - \Delta F}{T} 
\eeq
where by the first law of thermodynamics the 
change in the energy of the system is  
\beq
\Delta E  \ \ = \ \ \mathcal{W} - \mathcal{Q}
\eeq
The total entropy change of the universe is the sum 
of the system entropy change, and that of the bath 
\beq
\mathcal{S}  \ \ &=& \ \ \Delta S + \frac{\mathcal{Q}}{T}
\ \ = \ \ \frac{\mathcal{W}-\Delta F}{T}
\eeq
It follows that the Crooks relation can be written as 
\begin{equation}  
\frac{P_{0\leadsto1}(\mathcal{S}) }{P_{1\leadsto0}(-\mathcal{S})} 
\ \ = \ \ \exp\left[\mathcal{S}\right]
\end{equation}
As in the case of the Jarzynski equality we deduce that 
\beq
\Big\langle \exp\left[-\mathcal{S}\right] \Big\rangle 
\ \ = \ \ 1
\eeq
and consequently 
\beq
\left\langle \mathcal{S} \right\rangle \ \ > \ \ 0,
\eeq
in accordance with the second law of thermodynamics. 
Note that it is only the average $\langle\mathcal{S}\rangle$ that is positive. 
In a finite system $\mathcal{S}$ is negative for a fraction 
of the processes, with vanishing manifestation in the thermodynamic limit.

%%%%%%%%%%%%%%%%%%%%%%%%%%%%%%%%%%%%%%%%%%%%%%%%%%%%%%%%%%%%%%%%%%%%%%%%%%%%%%%%%%%%%%%%%%%%
\section{Summary}

The objective of this work was to illuminate that the simplicity 
of the NFT is maintained also in the quantum context. The way 
to go was to regard it a arising from a generalized detailed 
balance relation \Eq{CrooksE}. This connects smoothly 
with the formulation of the ``quantum fluctuation theorems for heat 
exchange in \cite{NFTx}.

A key issue was to regard the work agent as a dynamical entity, 
and to avoid a continuous measurement scheme for its measurement. 
This allowed us to bypass the subtlety that has been expressed 
in previous publications, such as \cite{CrooksJQ} that has been cited 
in the Introduction. If one would like to consider a multi-stage 
cycle in which the system interacts with several agents - there is no problem 
with that: the interaction with an agent has finite time duration, 
but once it is switched off we have an unlimited time to perform   
a projective measurement of the agent. Meanwhile the process protocol    
is not disturbed, and therefore a Markovian assumption is not 
required for the formulation, nor continuous measurement scheme. 

One may be troubled because the control parameters in our 
formulations become dynamical variables with quantum uncertainties.
However, this is hardly a criticism of our approach, since reality 
is in-fact quantum mechanical, hence this ``price" cannot be avoided.

It was also important to clarify the role of the environment.
Here a master equation approach might be illuminating, 
but it is not required in the derivation.  
In this context it was quite instructive to repeat 
the considerations in terms of the combined states of the system and the bath, 
in the manner suggested for example by Fano \cite{Fano} and Lipkin \cite{Lipkin}.  
\\

%%%%%%%%%%%%%%%%%%%%%%%%%%%%%%%%%%%%%%%%%%%%%%%%%%%%%%%%%%%%%%%%
{\bf Acknowledgments:}
The authors thank Michael Aizenman, Ariel Amir,  Yarden Cohen 
and Robert Dorfman for discussions.
This work was supported by the German Federal Ministry of
Education and Research (BMBF) within the framework of the
German-Israeli project cooperation (DIP), by the US-Israel
Binational Science Foundation (BSF), by the Israel Science
Foundation (ISF) and by its Converging Technologies Program.
Work by YI was partially supported by a continuing Humboldt 
Foundation research award.
%\end{acknowledgments}

%%%%%%%%%%%%%%%%%%%%%%%%%%%%%%%%%%%%%%%%%%%%%%%%%%%%%%%%%%%%%%%%

%%%%%%%%%%%%%%%%%%%%%%%%%%%%%%%%%%%%%%%%%%%%%%%%%%%%%%%%%%%%%%%%%%%%%%%%%%%%%%%%%%%%%%%%%%
\clearpage
\end{document}